# Estimating Information-Theoretic Quantities

Robin A.A. Ince[1], Simon R. Schultz[2] and Stefano Panzeri[1,3]

[1] *Institute of Neuroscience and Psychology, 58 Hillhead Street, University of Glasgow, Glasgow G12 8QB, UK*

[2] *Department of Bioengineering, Imperial College London, South Kensington, London SW7 2AZ, UK*

[3] *Center For Neuroscience and Cognitive Systems, Italian Institute of Technology, Corso Bettini 31 – 38068 Rovereto (Tn) Italy*

16 pages 5724 words.

## Definition

Information theory is a practical and theoretical framework developed for the study of communication over noisy channels. Its probabilistic basis and capacity to relate statistical structure to function make it ideally suited for studying information flow in the nervous system. It has a number of useful properties: it is a general measure sensitive to any relationship, not only linear effects; it has meaningful units which in many cases allow direct comparison between different experiments; and it can be used to study how much information can be gained by observing neural responses in single trials, rather than in averages over multiple trials. A variety of information theoretic quantities are in common use in neuroscience – (see entry "Summary of Information-Theoretic Quantities"). *Estimating* these quantities in an accurate and unbiased way from real neurophysiological data frequently presents challenges, which are explained in this entry.

# Detailed Description

## Information theoretic quantities

Information theory provides a powerful theoretical framework for studying communication in a quantitative way. Within the framework there are many different quantities to assess different properties of systems, many of which have been applied fruitfully to analysis of the nervous system and neural communication. These include the entropy (H; which measures the uncertainty associated with a stochastic variable), mutual information (I; which measures the dependence relationship between two stochastic variables), and several more general divergence measures for probability distributions (Kullbeck-Liebler, Jenson-Shannon). For full definitions and details of these quantities, see the "Summary of Information-Theoretic Quantities" entry.

## The Finite Sampling Problem

A major difficulty when applying techniques involving information theoretic quantities to experimental systems is that they require measurement of the full probability distributions of the variables involved. If we had an infinite amount of data, we could measure the true stimulus-response probabilities precisely. However, any real experiment only yields a finite number of trials from which these probabilities must be estimated. The estimated probabilities are subject to statistical error and necessarily fluctuate around their true values. The significance of these finite sampling fluctuations is that they lead to both statistical error (variance) and systematic error (called limited sampling bias) in estimates of entropies and information. This bias is the difference between the expected value of the quantity considered, computed from probability distributions estimated with N trials or samples, and its value computed from the true probability distribution. This is illustrated in Figure 1, which shows histogram estimates of the response distribution for two different stimuli calculated from 40 trials per stimulus. While the true probabilities are uniform (dotted black lines) the calculated maximum likelihood estimates show some variation around the true values. This causes spurious differences – in this example the sampled probabilities indicate that obtaining a large response suggests stimulus 1 is presented – which results in a positive value of the information. Figure 1C shows a histogram of information values obtained from many simulations of this system. The bias constitutes a significant practical problem, because its magnitude is often of the order of the information values to be evaluated, and because it cannot be alleviated simply by averaging over many neurons with similar characteristics.

### Direct estimation of information theoretic quantities

The most direct way to compute information and entropies is to estimate the response probabilities as the experimental histogram of the frequency of each response across the available trials, and then plug these empirical probability estimates into Eqs. (1-3). We refer to this as the "plug-in" method.

The plug-in estimate of entropy is biased downwards (estimated values are lower than the true value) while the plug-in estimate of information shows an upward bias. An intuitive explanation for why entropy is biased downwards comes from the fact that entropy is a measure of variability. As we saw in Figure 1, variance in the maximum likelihood probability estimates introduces extra structure in the probability distributions; this additional variation (which is dependent on the number of trials from the asymptotic normality of the ML estimate) always serves to make the estimated distribution less uniform or structured than the true distribution. Consequently, entropy estimates are lower than their true values, and the effect of finite sampling on entropies is a downward bias. For information, an explanation for why bias is typically positive is that finite sampling can introduce spurious stimulus-dependent differences in the response probabilities, which make the stimuli seem more discernible and hence the neuron more informative than it really is. Alternatively, viewing the information as a difference of unconditional and conditional entropies, (first two expressions of Eq. in "Summary of Information Theoretic Quantities") the conditional entropy for each stimulus value is estimated from a reduced number of trials compared to the noise entropy (only the trials corresponding to that stimulus) and so its bias is considerably larger. Since entropy is biased down and the conditional entropy term appears with a negative sign, this causes the upward bias in information.

In the above explanations for the source of the bias, the source is the variability in estimates of the probability distributions considered. Variance in the maximum likelihood histogram estimators of the probabilities induces additional "noise" structure affecting the entropy and information values.

### Asymptotic estimates of the limited sampling bias

To understand the sampling behaviour of information and entropy better, it is useful to find analytical approximations to the bias. This can be done in the so-called "asymptotic sampling regime". Roughly speaking this is when the number of trials is "large". More rigorously, the asymptotic sampling regime is defined as $N$ being large enough that every possible response occurs many times: that is, $N_s P(r|s) \gg 1$ for each stimulus-response pair $s,r$ such that $P(r|s) > 0$. In this regime, the bias of the entropies and information can be expanded in inverse powers of $1/N$ and analytical approximations obtained (Miller, 1955; Panzeri and Treves, 1996). The leading terms in the biases are respectively:

$$BIAS[H(R)] = \frac{-1}{2N \ln(2)} \left[ \overline{R} - 1 \right]$$

$$BIAS[H(R|S)] = \frac{-1}{2N \ln(2)} \sum_s \left[ \overline{R}_s - 1 \right]$$

$$BIAS[I(S;R)] = \frac{1}{2N \ln(2)} \left\{ \sum_s \left[ \overline{R}_s - 1 \right] - \left[ \overline{R} - 1 \right] \right\}$$

(1)

Here $\overline{R}_s$ denotes the number of relevant responses for the stimulus conditional response probability distribution $P(r|s)$ – i.e. the number of different responses $r$ with non-zero probability of being observed when stimulus s is presented. $\overline{R}$ denotes the number of relevant responses for $P(r)$ – i.e. the number of different responses $r$ with non-zero probability of being observed across all stimuli. In practice, $\overline{R}$ is usually going to be equal to the number of elements constituting the response space $R$. If a response never happens across all stimuli, it can be removed from the sum over r in the calculation of each entropy term, and thus removed from the response set $R$. However, $\overline{R}_s$ may be different from $\overline{R}$ if some responses occur only with particular stimuli.

Although valid only in the asymptotic regime, Eq. (1) sheds valuable light on the key factors that control the bias. First, Eq. (1) shows that the bias of $H(R|S)$ is approximately $S$ times bigger than that of $H(R)$. This means that, in the presence of many stimuli, the bias of $I(S;R)$ is similar to that of $H(R|S)$. However, $I(S;R)$ is a difference of entropies, and its typical values are much smaller than those of $H(R|S)$. This implies that bias correction methods must be validated on the performance of information and not only on entropies, because, in many cases, the bias may be proportionally negligible for entropies but not the information. Second, Eq. (1) shows that the bias is small when the ratio $N_s/\overline{R}$ is big, *i.e.* more trials per stimulus than possible responses. This is because, assuming that the number of trials per stimulus is approximately constant and $\overline{R}_s$ is approximately equal to $\overline{R}$, the bias of $H(R|S)$ is approximately $-\overline{R}$ /[$2N_s\ln(2)$]. Thus, $N_s/\overline{R}$ is *the* crucial parameter for the sampling problem. For example, in the simulations of Fig. 2, with $\overline{R}=2^8$ the bias of $I(S;R)$ became negligible for $N_s \geq 2^{13}$ (i.e. $N_s/\overline{R} \geq 32$). Second, Eq. (1) shows that, even if $N_s$ is constant, the bias increases with the number of responses $\overline{R}$. This has important implications for comparing neural codes. For example, a response consisting of a given number of spikes can arise from many different possible temporal patterns of spikes. Thus, $\overline{R}$ is typically much greater for a spike timing code than for a spike count code, and it follows from Eq. (1) that the estimate of information conveyed by a spike timing code is more biased than that measured for the same neurons through a spike count code. If bias is not eliminated, there is therefore a danger of concluding that spike timing is important even when it is not.

A further important feature of Eq. (1) is that, although the bias is not the same for all probability distributions, in the asymptotic sampling regime it depends on some remarkably simple details of the response distribution (the number of trials and the number of relevant responses). Thus Eq. (1) makes it is possible to derive simple rules of thumb for estimating the bias magnitude and compare the relative bias in different situations. As detailed in the next section, the simplicity of Eq. (1) can also be exploited in order to correct effectively for the bias (Panzeri and Treves, 1996).

## Bias Correction Methods

A number of techniques have been developed to address the issue of bias, and allow much more accurate estimates of information theoretic quantities than the "plug-in" method described above. (Panzeri et al., 2007) and (Victor, 2006) provide a review of such methods, a selection of which are briefly outlined here.

### Panzeri–Treves (PT)

In the so-called asymptotic sampling regime, when the number of trials is large enough that every possible response occurs many times, an analytical approximation for the bias (i.e. the difference between the true value and the plug-in estimate) of entropies and information can be obtained (Miller, 1955; Panzeri and Treves, 1996) (Eq. (1)). The value of the bias computed from the above expressions is then subtracted from the plug-in estimate to obtain the corrected value. This requires an estimate of the number of relevant responses for each stimulus, $\overline{R}_s$. The simplest approach is to approximate $\overline{R}_s$ by the count of responses that are observed at least once – this is the "naive" count. However due to finite sampling this will be an underestimate of the true value. A Bayesian procedure (Panzeri and Treves, 1996) can be used to obtain a more accurate value.

### Quadratic Extrapolation (QE)

In the asymptotic sampling regime, the bias of entropies and information can be approximated as second order expansions in 1/N, where N is the number of trials (Treves and Panzeri, 1995; Strong et al., 1998). For example, for information:

$$I_{plugin}(R;S) = I_{true}(R;S) + \frac{a}{N} + \frac{b}{N^2}$$

(2)

This property can be exploited by calculating the estimates with subsets of the original data, with N/2 and N/4 trials and fitting the resulting values to the polynomial expression above. This allows an estimate of the parameters *a* and *b* and hence $I_{true}(S; R)$. To use all available data, estimates of two subsets of size N/2 and four subsets of size N/4 are averaged to obtain the values for the extrapolation. Together with the full length data calculation, this requires seven different evaluations of the quantity being estimated.

### Nemenman–Shafee–Bialek (NSB)

The NSB method (Nemenman et al., 2002, 2004) utilises a Bayesian inference approach and does not rely on the assumption of the asymptotic sampling regime. It is based on the principle that when estimating a quantity, the least bias will be achieved when assuming an a priori uniform distribution over the quantity. This method is more challenging to implement than the other methods, involving a large amount of function inversion and numerical integration. However, it often gives a significant improvement in the accuracy of the bias correction (Montani et al., 2007, 2007; Montemurro et al., 2007; Nemenman et al., 2008).

### Shuffled Information Estimator ($I_{sh}$)

Recently, an alternative method of estimating the mutual information has been proposed (Montemurro et al., 2007; Panzeri et al., 2007). Unlike the methods above, this is a method for calculating the information only, and is not a general entropy bias correction. However, it can be used with the entropy corrections described above to obtain more accurate results. In brief, the method uses the addition and subtraction of different terms (in which correlations given a fixed stimulus are removed either by empirical shuffling or by marginalizing and then multiplying response probabilities) that do not change the information value in the limit of an infinite number of trials, but do remove a large part of the bias for finite number of trials. This greatly reduces the bias of the information estimate, at the price of only a marginal increase of variance. Note that this procedure works well for weakly correlated data, which is frequently the case for simultaneously recorded neurons.

### Examples of bias correction performance

Figure 2A reports the results of the performance of bias correction procedures, both in terms of bias and Root Mean Square (RMS) error, on a set of simulated spike trains from eight simulated neurons. Each of these neurons emitted spikes with a probability obtained from a Bernoulli process. The spiking probabilities were set equal to those measured, in the 10–15 ms post-stimulus interval, from eight neurons in rat somatosensory cortex responding to 13 stimuli consisting of whisker vibrations of different amplitude and frequency (Arabzadeh et al., 2004). The 10–15 ms interval was chosen since it was found to be the interval containing highest information values. Figure 2A shows that all bias correction procedures generally improve the estimate of *I(S;R)* with respect to the plug-in estimator, and the NSB correction is especially effective.

Figure 2B shows that the bias-corrected estimation of information (and RMS error; lower panel) is much improved by using $I_{sh}(R;S)$ rather than *I(R;S)*. The use of $I_{sh}(R;S)$ makes the residual errors in the estimation of information much smaller and almost independent from the bias correction method used. Taking into account both bias correction performance and computation time, for this simulated system the best method to use is the shuffled information estimator combined with the Panzeri–Treves analytical correction. Using this, an accurate estimate of the information is possible even when the number of samples per stimulus is R where R is the dimension of the response space.

It should be noted that while bias correction techniques such as those discussed above are essential if accurate estimation of the information is required, whether to employ them depends upon the question being addressed. If the aim is quantitative comparison of information values between different experiments, stimuli or behaviours, they are essential. If instead the goal is simply to detect the presence of a significant interaction, the statistical power of information as a test of independence is greater if the uncorrected plug-in estimate is used (Ince et al., 2012). This is because all the bias correction techniques introduce additional variance; for a hypothesis test which is unaffected by underlying bias, this variance reduces the statistical power.

## Information Component Analysis Techniques

The principal use of Information Theory in neuroscience is to study the neural code – and one of the commonly studied problems in neural coding is to take a given neurophysiological dataset, and to determine how the information about some external correlate is represented. One way to do this is to compute the mutual information that the neural responses convey on average about the stimuli (or other external correlate), under several different assumptions about the nature of the underlying code (e.g. firing rate, spike latency, etc). The assumption is that if one code yields substantially greater mutual information, then downstream detectors are more likely to be using that code (the *Efficient Coding Hypothesis*).

A conceptual improvement on this procedure, where it can be done, is to take the total information available from all patterns of spikes (and silences), and to break it down (mathematically) into components reflecting different encoding mechanisms, such as firing rates, pairwise correlations, etc (Panzeri et al., 1999; Panzeri and Schultz, 2001). One way to do this is to perform an asymptotic expansion of the mutual information, grouping terms in such a way that they reflect meaningful coding mechanisms.

### Taylor series expansion of the mutual information

A taylor series expansion of the mutual information is one such approach. For short time windows (and we will discuss presently what "short" means), the mutual information can be approximated as

$$I(R;S) = TI_t + \frac{T^2}{2} I_{tt} + ... \qquad (3)$$

where the subscript *t* indicates the derivative with respect to the time window length *T*. For time windows sufficiently short that only a pair of spikes are contained within the time window (either from a population of cells or from a single spike train), the first two terms are all that is needed. If the number of spikes exceeds two, it may still be a good approximation, but higher order terms would be needed to capture all of the information. One possibility is that Equation (3) could be extended to incorporate higher order terms, however we have instead found it better in practice to take an alternative approach (see next section).

With only a few spikes to deal with, it is possible to use combinatorics to write out the expressions for the probabilities of observing different spike patterns. This was initially done for the information contained in the spikes fired by a small population of cells (Panzeri et al., 1999), and then later extended to the information carried by the spatiotemporal dynamics of a small population of neurons over a finite temporal wordlength (Panzeri and Schultz, 2001). In the former formulation, we define $\bar{r}_i(s)$ to be the mean response rate (number of spikes in *T* divided by *T*) of cell *i* (of *C* cells) to stimulus *s* over all trials where *s* was presented. We then define the signal cross-correlation density to be

$$v_{ij} = \frac{\langle \bar{r}_i(s)\bar{r}_j(s)\rangle_s}{\langle \bar{r}_i(s)\rangle_s \langle \bar{r}_j(s)\rangle_s} - 1 \quad . \tag{4}$$

This quantity captures the correlation in the mean response profiles of each cell to different stimuli (e.g. correlated tuning curves). Analogously, we define the noise cross-correlation density to be

$$\gamma_{ij}(s) = \frac{\overline{r_i(s)r_j(s)}}{\overline{r_i(s)}\,\overline{r_j(s)}} - 1 \quad . \tag{5}$$

The mutual information *I(R;S)* can then be written as the sum of four components,

$$I = I_{\text{lin}} + I_{\text{sig-sim}} + I_{\text{cor-ind}} + I_{\text{cor-dep}} \quad , \tag{6}$$

with the first component

$$I_{\text{lin}} = T\sum_{i=1}^{C}\left\langle \bar{r}_i(s)\log_2 \frac{\bar{r}_i(s)}{\langle \bar{r}_i(s')\rangle_{s'}}\right\rangle_s \tag{7}$$

capturing the rate component of the information, i.e. that which survives when there is no correlation between cells (even of their tuning curves) present at all. This quantity is positive semi-definite, and adds linearly across neurons. It is identical to the "information per spike" approximation calculated by a number of authors (Skaggs et al., 1993; Brenner et al., 2000; Sharpee et al., 2004). $I_{\text{sig-sim}}$ is the correction to the information required to take account of correlation in the tuning of individual neurons, or *signal similarity*:

$$I_{\text{sig-sim}} = \frac{T^2}{2\ln 2}\sum_{i=1}^{C}\sum_{j=1}^{C}\langle \bar{r}_i(s)\rangle_s \langle \bar{r}_j(s)\rangle_s \left[v_{ij} + (1+v_{ij})\ln\left(\frac{1}{1+v_{ij}}\right)\right] \tag{8}$$

This is negative semi-definite. $I_{\text{cor-ind}}$ is the effect on the transmitted information of the average level of noise correlation (correlation at fixed stimulus) between neurons:

$$I_{\text{cor-ind}} = \frac{T^2}{2}\sum_{i=1}^{C}\sum_{j=1}^{C}\langle \bar{r}_i(s)\bar{r}_j(s)\gamma_{ij}(s)\rangle_s \log_2\left(\frac{1}{1+v_{ij}}\right) . \tag{9}$$

$I_{\text{cor-ind}}$ can take either positive or negative values. $I_{\text{cor-dep}}$ is the contribution of stimulus-dependence in the correlation to the information – as would be present, for instance, if synchrony was modulated by a stimulus parameter:

$$I_{\text{cor-dep}} = \frac{T^2}{2}\sum_{i=1}^{C}\sum_{j=1}^{C}\left\langle \bar{r}_i(s)\bar{r}_j(s)(1+\gamma_{ij}(s))\log_2\left[\frac{\langle \bar{r}_i(s')\bar{r}_j(s')\rangle_{s'}(1+\gamma_{ij}(s))}{\langle \bar{r}_i(s')\bar{r}_j(s')(1+\gamma_{ij}(s'))\rangle_{s'}}\right]\right\rangle_s \tag{10}$$

An application of this approach to break out components reflecting rate and correlational coding is illustrated in Figure 3, by application to simulated spike trains with controlled correlation. This approach has been used by a number of authors to dissect out contributions to neural coding (e.g. Scaglione et al., 2008).

This approach extends naturally to the consideration of temporal correlations between spike times within a spike train (Panzeri and Schultz, 2001). Note that the Taylor series approach can be applied to the entropy as well as the mutual information – this was used to break out the effect of spatiotemporal correlations on the entropy of a small population of neural spike trains (Schultz and Panzeri, 2001). While the Taylor series approach can be extremely useful in teasing out contributions of different mechanisms to the transmission of information, it is not recommended as a method for estimation of the total information, as the validity of the approximation can break down quickly as the time window grows beyond that sufficient to contain more than a few spikes from the population. It is however useful as an additional inspection tool after the total information has been computed, which allows the information to be broken down into not only mechanistic components but also their contributions from individual cells and time bins (for instance by visualizing the quantity after the summations in Equation (10) as a matrix).

## A more general information component analysis approach

A limitation of the Taylor series approach is that it is restricted to the analysis of quite short time windows of neural responses. However, an exact breakdown approach allowed a very similar information component approach to be used (Pola et al., 2003). In this approach, we consider a population response (spike pattern) *r*, and define the normalized noise cross-correlation to be

$$\gamma(\mathbf{r}|s) = \begin{cases} \dfrac{P(\mathbf{r}|s)}{P_{ind}(\mathbf{r}|s)} - 1 & \text{if } P_{ind}(\mathbf{r}|s) \neq 0 \\ 0 & \text{if } P_{ind}(\mathbf{r}|s) = 0 \end{cases}. \tag{11}$$

where the marginal distribution

$$P_{ind}(\mathbf{r}|s) = \prod_{c=1}^{C} P(r_c|s) . \tag{12}$$

Compare the expressions in Equation (5) and (12). Similarly, the signal correlation becomes

$$v(\mathbf{r}) = \begin{cases} \dfrac{P_{ind}(\mathbf{r})}{\prod_c P(r_c)} - 1 & \text{if } \prod_c P(r_c) \neq 0 \\ 0 & \text{if } \prod_c P(r_c) = 0 \end{cases}. \tag{13}$$

Using these definitions, the information components can now be written exactly (without the approximation of short time windows) as

$$I_{lin} = \sum_c \sum_{r_c} \left\langle P(r_c|s) \log_2 \dfrac{P(r_c|s)}{P(r_c)} \right\rangle_s \tag{14}$$

$$I_{\text{sig-sim}} = \frac{1}{\ln 2} \sum_{\mathbf{r}} \left( \prod_c P(r_c) \right) \left[ \nu(\mathbf{r}) + (1+\nu(\mathbf{r})) \ln\left( \frac{1}{1+\nu(\mathbf{r})} \right) \right] \quad (15)$$

$$I_{\text{cor-ind}} = \sum_{\mathbf{r}} \left\langle P_{\text{ind}}(\mathbf{r}|s) \gamma(\mathbf{r}|s) \right\rangle_s \log_2 \left( \frac{1}{1+\nu(\mathbf{r})} \right) \quad (16)$$

$$I_{\text{cor-dep}} = \sum_{\mathbf{r}} \left\langle P_{\text{ind}}(\mathbf{r}|s)(1+\gamma(\mathbf{r}|s)) \log_2 \frac{\left\langle P_{\text{ind}}(\mathbf{r}|s') \right\rangle_{s'} (1+\gamma(\mathbf{r}|s))}{\left\langle P_{\text{ind}}(\mathbf{r}|s')(1+\gamma(\mathbf{r}|s')) \right\rangle_{s'}} \right\rangle_s . \quad (17)$$

This latter component is identical to the quantity "$\Delta I$" introduced by Latham and colleagues (Nirenberg et al., 2001) to describe the information lost due to a decoder ignoring correlations. It is zero if and only if $P(s|\mathbf{r}) = P_{\text{ind}}(s|\mathbf{r})$ for every $s$ and $\mathbf{r}$.

The expressions shown above are perhaps the most useful description for obtaining insight into the behavior of the information components under different statistical assumptions relating to the neural code, however they are not necessarily the best way to *estimate* the individual components. However, the components can also be written as a sum of entropies and entropy-like quantities, which can then be computed using the entropy estimation algorithms described earlier in this chapter (Pola et al., 2003; Montani et al., 2007; Schultz et al., 2009). Note that, as shown by Scaglione and colleagues (Scaglione et al., 2008, 2010) the components in Eq. (14-17) can, under appropriate conditions, be further decomposed to tease apart the relative role of autocorrelations (spikes from the same cells) and cross-correlations (spikes from different cells).

**Maximum entropy approach to information component analysis**

The conceptual basis of the information component approach is to make an assumption that constrains the statistics of the response distribution $P(r|s)$, compute the mutual information subject to this assumption, and by evaluating the difference between this and the "full" information, calculate the contribution to the information of relaxing this constraint. By making further assumptions, the contributions of additional mechanisms can in many cases be dissected out hierarchically. As an example, by assuming that the conditional distribution of responses given stimuli is equal to the marginal distribution $P(\mathbf{r}|s) = P_{\text{ind}}(\mathbf{r}|s)$, and substituting in to the mutual information equation, one can define an information component $I_{ind}$. This component can then be further broken up

$$I_{\text{ind}} = I_{\text{lin}} + I_{\text{sig-sim}}, \quad (18)$$

with $I_{\text{lin}}$ and $I_{\text{sig-sim}}$ as defined in the previous section. The correlational component, $I_{cor}$, is then just the difference between $I_{\text{ind}}$ and the total mutual information. This can also be further broken up, as

$$I_{\text{cor}} = I_{\text{cor-ind}} + I_{\text{cor-dep}}. \quad (19)$$

This approach can be extended further. For instance, the assumption of a Markov approximation can be made, with a memory extending back $q$ timesteps, and the information computed under this assumption (Pola et al., 2005). More generally, any simplified model can be used, although the maximum entropy models are of special interest (Montemurro et al., 2007; Schaub and Schultz, 2012)

$$P_{\text{simp}}(\mathbf{r} | s) = \frac{1}{Z} \exp\left\{\lambda_0 - 1 + \sum_{i=1}^{m} \lambda_i g_i(\mathbf{r})\right\} \quad (20)$$

with parameters $\lambda_i$ implementing a set of constraints reflecting assumptions made about what are the important (non-noise) properties of the empirical distribution, and the partition function Z ensuring normalisation. An example of this is the Ising model, which has been used with some success to model neural population response distributions (Schneidman et al., 2006; Shlens et al., 2006; Schaub and Schultz, 2012). Estimation of the Ising model for large neural populations can be extremely computationally intensive if brute force methods for computing the partition function are employed, however mean field approximations can be employed in practice with good results (Roudi et al., 2009; Schaub and Schultz, 2012).

## Binless methods for estimating information

In applications in neuroscience, typically at least one of the random variables (stimuli or responses) is discrete, and thus the approach of discretizing one or more of the variables is often taken. However, where stimuli and responses are both continuous (an example might be local field potential responses to a white noise sensory stimulus), it may be advantageous to take advantages of techniques better suited to continuous signals, such as kernel density estimators (Moon et al., 1995), nearest neighbor estimators (Kraskov et al., 2004) or *binless* metric space methods (Victor, 2002). We refer to the entry "Bin-Less Estimators for Information Quantities" for an in-depth discussion of these techniques.

## Calculation of information theoretic quantities from parametric models

An alternative to measuring information-theoretic quantities directly from observed data is to fit the data to a model, and then to either analytically or numerically (depending upon the complexity of the model) compute the information quantity from the model. Examples include analytical calculations of the information flow through models of neural circuits with parameters fit to anatomical and physiological data (Treves and Panzeri, 1995; Schultz and Treves, 1998; Schultz and Rolls, 1999), and parametric fitting of probabilistic models of neural spike firing (Paninski, 2004; Pillow et al., 2005, 2008), of spike count distributions (Gershon et al., 1998; Clifford and Ibbotson, 2000), or of neural mass signals (Magri et al., 2009). It should come as no surprise that "there is no free lunch", and, although in principle information quantities can be computed exactly under such circumstances, the problem is moved to one of assessing model validity – an incorrect model can lead to a bias in either direction in the information computed.

## Acknowledgements

Research supported by the SI-CODE (FET-Open, FP7-284533) project and by the ABC and NETT (People Programme Marie Curie Actions PITN-GA-2011-290011 and PITN-GA-2011-289146) projects of the European Union's Seventh Framework Programme FP7 2007-2013.

# Figures and Figure Captions

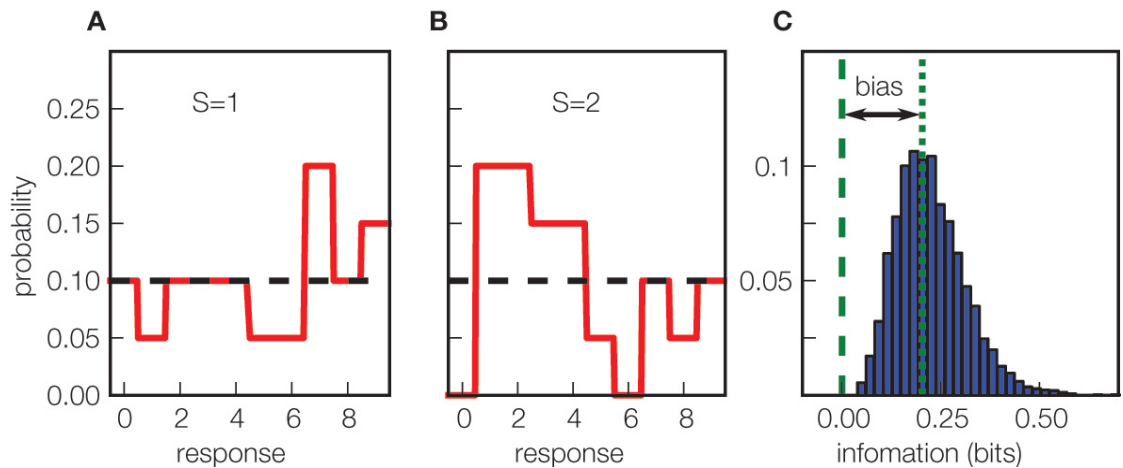

**Figure 1:** *The origin of the limited sampling bias in information measures.* (**A**, **B**) Simulation of a toy uninformative neuron, responding on each trial with a uniform distribution of spike counts ranging from 0 to 9, regardless of which of two stimuli (S = 1 in (A) and S = 2 in (B)) are presented. The black dotted horizontal line is the true response distribution, solid red lines are estimates sampled from 40 trials. The limited sampling causes the appearance of spurious differences in the two estimated conditional response distributions, leading to an artificial positive value of mutual information. (**C**) The distribution (over 5000 simulations) of the mutual information values obtained (without using any bias correction) estimating Eq. 1 from the stimulus–response probabilities computed with 40 trials. The dashed green vertical line indicates the true value of the mutual information carried by the simulated system (which equals 0 bits); the difference between this and the mean observed value (dotted green line) is the bias. Redrawn with permission from (Ince et al., 2010).

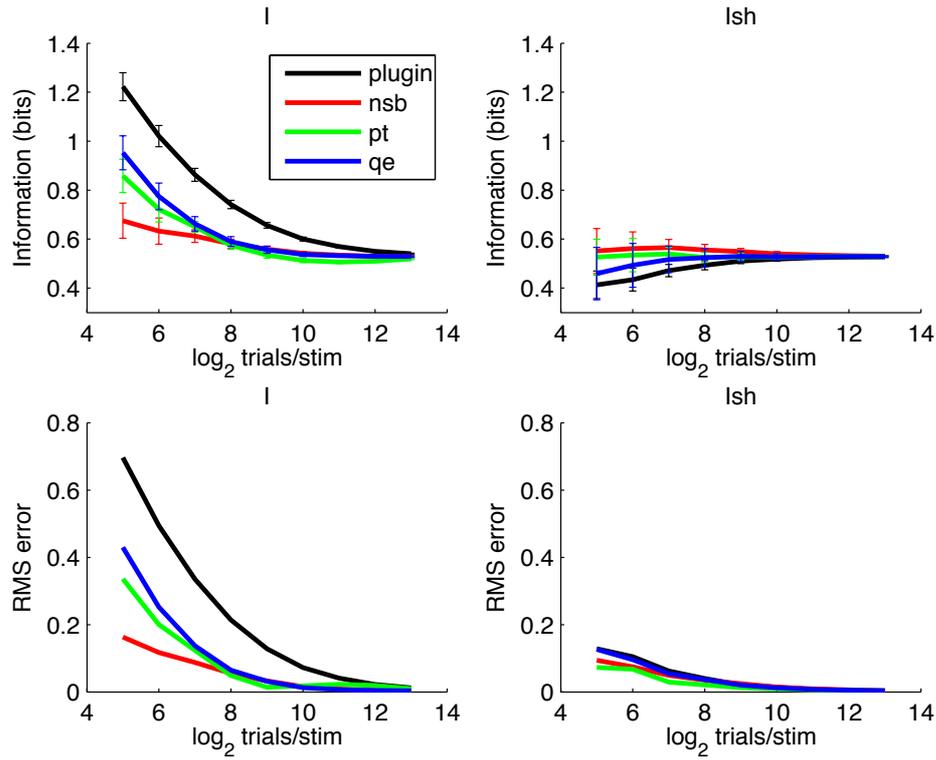

**Figure 2:** *Performance of various bias correction methods.* Several bias correction methods were applied to spike trains from eight simulated somatosensory cortical neurons. The information estimates (upper panels) and root mean square (RMS) error (lower panels) are plotted as a function of the simulated number of trials per stimulus. (**A**) Mean +/- SD (upper panel; over 50 simulations) and RMS error (lower panel) of *I(S;R)*. (**B**) Mean +/- SD (upper panel; over 50 simulations) and RMS error (lower panel) of $I_{sh}(S;R)$.

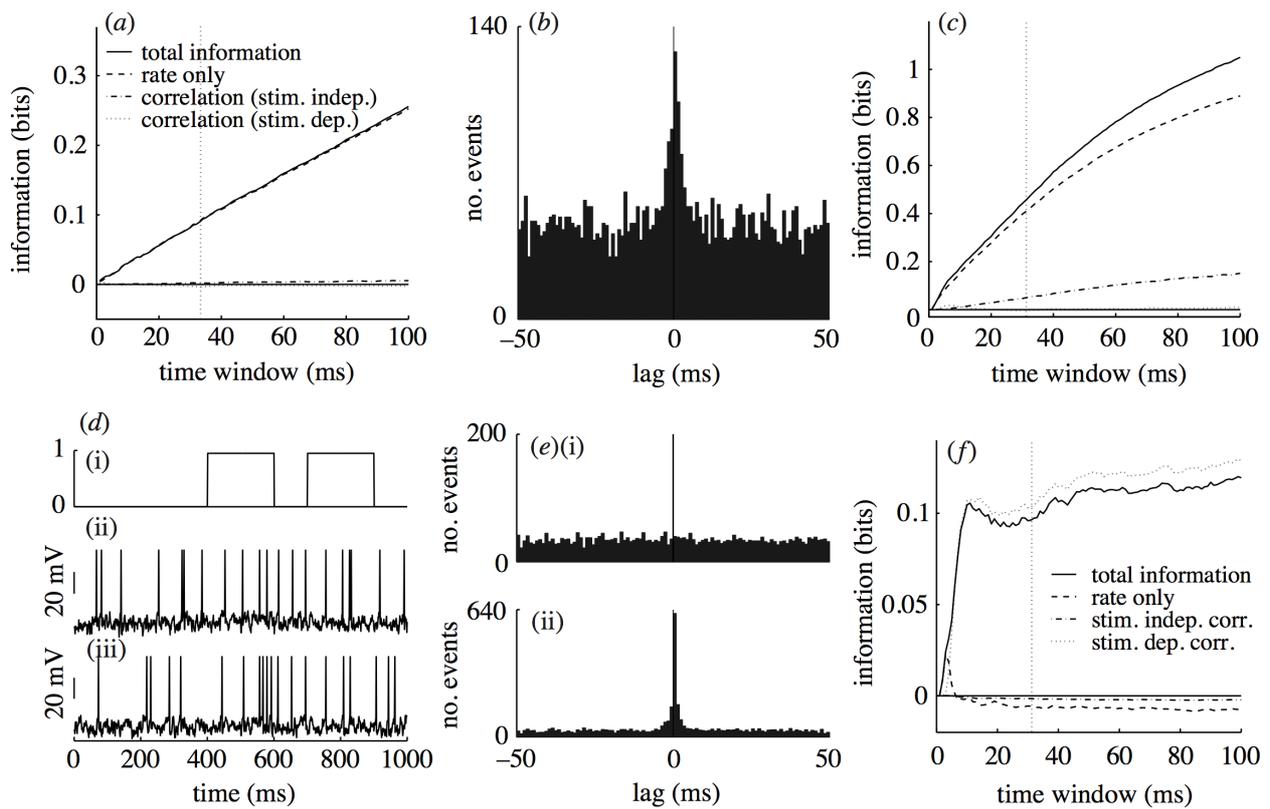

**Figure 3.** Information component analysis of simulated data (for a five-cell ensemble). (a) Poisson cells, with the only difference between the stimuli being firing rate. (b) Integrate-and-fire cells, with common input due to sharing of one third of connections, resulting in a cross-correlogram as depicted, leads to (c) contribution to the information from $I_{cor\text{-}ind}$. (d) Integrate-and-fire cells with two of the five cells increasing their correlation due to increased shared input, for one of the stimuli, where the others remain randomly correlated, leading to cross-correlograms shown in panel (e). Firing rates are approximately balanced between the two stimuli. This results in a substantial contribution to the information from the stimulus-dependent correlation component, $I_{cor\text{-}dep}$. From (Panzeri et al., 1999) with permission.